\documentclass[twocolumn,letter]{jpsj2} 
\title{Evidence for Novel Pairing State in Noncentrosymmetric Superconductor CePt$_{3}$Si: $^{29}$Si-NMR Knight Shift Study}

\author{
Mamoru \textsc{Yogi}$^{1}$\thanks{E-mail: myogi@sci.u-ryukyu.ac.jp}\thanks{Present address: Faculty of Science, University of the Ryukyus, Okinawa 903-0213},
Hidekazu \textsc{Mukuda}$^{1}$,
Yoshio \textsc{Kitaoka}$^{1}$,
Shin \textsc{Hashimoto}$^{2}$,
Takashi \textsc{Yasuda}$^{2}$,
Rikio \textsc{Settai}$^{2}$,
Tatsuma D \textsc{Matsuda}$^{3}$,
Yoshinori \textsc{Haga}$^{3}$,
Yoshichika \textsc{\={O}nuki}$^{2,3}$,
Peter \textsc{Rogl}$^{4}$
and Ernst \textsc{Bauer}$^{5}$
}

\inst{
$^1$Department of Materials Engineering Science, Graduate School of Engineering Science, Osaka University,  Osaka 560-8531, Japan\\
$^2$Graduate School of Science, Osaka University, Toyonaka, Osaka 560-0043, Japan\\
$^3$Advanced Science Research Center, Japan Atomic Energy Research Institute, Tokai, Ibaraki 319-1195, Japan\\
$^4$Institut f\"{u}r Physikalische Chemie, Universit\"{a}t Wien, A-1090 Wien, Austria\\
$^5$Institut f\"{u}r Festk\"{o}rperphysik, Technische Universit\"{a}t Wien, A-1040 Wien, Austria
}

\abst{
We report the measurements of the $^{29}$Si Knight shift $^{29}K$ on the noncentrosymmetric heavy-fermion compound CePt$_{3}$Si in which antiferromagnetism (AFM) with $T_{\rm N}=2.2$ K coexists with superconductivity (SC) with $T_{c}=0.75$ K. Its spin part $^{29}K_{\rm s}$, which is deduced to be $K_{\rm s}^{c}\ge 0.11$ and $0.16$\% at respective magnetic fields $H=2.0061$ and $0.8671$ T, does not decrease across the superconducting transition temperature $T_{c}$ for the field along the $c$-axis.  The temperature dependence of nuclear spin-lattice relaxation of $^{195}$Pt below $T_{c}$ has been accounted for by a Cooper pairing model with a two-component order parameter composed of spin-singlet and spin-triplet pairing components. From this result, it is shown that the Knight-shift data are consistent with the occurrence of the two-component order parameter for CePt$_{3}$Si.
}

\kword{heavy-fermion superconductivity, CePt$_{3}$Si, NMR, Knight shift, Cooper pairing symmetry}

\begin{document}
\maketitle
\newpage

The recent discovery of superconductivity (SC) in the noncentrosymmetric heavy-fermion (HF) compound CePt$_3$Si has initiated much interest \cite{CePt3Si_Bauer}, because experimental data revealed various intriguing features.\cite{CePt3Si_Bonalde,CePt3Si_Yogi,CePt3Si_Amato,CePt3Si_Metoki, CePt3Si_Izawa}
One of the novel superconducting characteristics of CePt$_{3}$Si is that the upper critical field $H_{c2}$ is significantly large, suggesting an extrapolated value of $H_{c2}\sim 3-5$ T, 3-5 times larger than the Pauli limit of $H_{c2}$ estimated from the relation $H_{\rm P}\approx \Delta/\sqrt{2}\mu_{\rm B}\approx 1$ T. 
A recent theoretical calculation revealed that such an enhancement of $H_{c2}$ is characteristic for the noncentrosymmetric SC in which the Cooper pairs are predominant in spin-triplet configurations.\cite{CePt3Si_Frigeri1} 
Another novel aspect is that a possible formation of spin-triplet pairing is realized under the uniform coexistence of antiferromagnetism (AFM) and SC, which was confirmed from the measurements of $^{195}$Pt nuclear spin-lattice relaxation rate $1/T_{1}$, muon spin rotation, and neutron scattering.\cite{CePt3Si_Yogi, CePt3Si_Amato, CePt3Si_Metoki}

With respect to the superconducting characteristics, most remarkably, the results of $^{195}$Pt $1/T_{1}$\cite{CePt3Si_Yogi}, have revealed that CePt$_3$Si is the first HF superconductor that exhibits a peak in $1/T_1$ just below $T_c$ and it does not follow the $T^3$ law for most unconventional HF superconductors.\cite{NMR_Tcube1, NMR_Tcube2} This unexpected  nature of SC is related to the absence of an inversion center in its crystal structure. To account for the experimentally observed features of the peak in $1/T_1$ just below $T_{c}$ \cite{CePt3Si_Yogi} and a line-node gap behavior at low temperatures detected by the measurements of London penetration depth \cite{CePt3Si_Bonalde} and thermal conductivity,\cite{CePt3Si_Izawa} a novel Cooper pairing model has been developed by assuming a two-component order parameter composed of spin-singlet and spin-triplet pairing components.\cite{CePt3Si_Hayashi,CePt3Si_Fujimoto} 
In particular,  Hayashi {\it et al.} have demonstrated that such a model accounts for the unique $T$ dependence of $1/T_1$ below $T_{c}$ on a qualitative level.\cite{CePt3Si_Hayashi}
To gain further insight into the novel order parameter symmetry of this compound, it is necessary to measure the spin susceptibility $\chi_{\rm s}$ below $T_{c}$ to distinguish between spin-singlet pairing from triplet pairing.

In materials without spatial inversion symmetry, the spin degeneracy of conduction electrons can be lifted by an antisymmetric spin-orbit coupling. 
The influence of this spin-orbit coupling on $\chi_{\rm s}$ has been treated theoretically on such superconductors, with a particular emphasis on the HF SC CePt$_3$Si.\cite{CePt3Si_Frigeri2,CePt3Si_Samokhin2} 
It was found that for this compound forming in the tetragonal crystal symmetry, irrespective of the pairing symmetry, the stable superconducting phases would give a very weak change in $\chi_{\rm s}$ for fields along the $c$-axis and a moderate reduction for fields in the basal plane. 
Frigeri {\it et al.} predicted that a spin-triplet state is not affected by the spin-orbit coupling, and the difference in $\chi_{\rm s}$ between spin-singlet and spin-triplet becomes smaller with increasing spin-orbit coupling strength. \cite{CePt3Si_Frigeri2}
Furthermore, Samokhin predicted that CePt$_{3}$Si has line nodes and the characteristic $T$ dependence of $\chi_{\rm s}$ such as $\chi_{\rm s}\propto T$ for the fields in the basal plane, and $\chi_{\rm s}\propto T^3$ for the fields along the $c$-axis \cite{CePt3Si_Samokhin2}.

$\chi_{\rm s}$ is measured using the NMR Knight shift ($K$) in superconductors to distinguish between spin-singlet and spin-triplet pairing.
The $\chi_{\rm s}$ for various HF superconductors was discussed on the basis of the experimental results of $K$, $1/T_1$, and  electronic specific heat ($\gamma_{el}$) within the framework of the Fermi liquid model for a Kramers doublet crystal-electric-field (CEF) ground state.\cite{NMR_Ks_Tou}
$\chi_{\gamma}$ was calculated from the enhanced Sommerfeld coefficient $\gamma_{el}$, and $\chi_{T_1}$ from the quasiparticle Korringa relation $T_{1}T(K_{T_1})^2={\rm const.}$ via the relation $\chi_{T_{1}}=(N_{\rm A}\mu_{\rm B}/A_{\rm hf})K_{T_1}$, where $A_{\rm hf}$ is the hyperfine coupling constant, $N_{\rm A}$ is Avogadoro's number and $\mu_{\rm B}$ is the Bohr magneton.
For the even-parity (spin-singlet) superconductors CeCu$_2$Si$_2$, CeCoIn$_5$ and UPd$_2$Al$_3$, the decrease in $K(T)$ below $T_{c}$ is caused by the decrease in $\chi_{\rm s}$ of the heavy quasiparticle estimated consistently from $\chi_{\gamma}$ and $\chi_{T_1}$.
This result indicates that heavy quasiparticles form the spin-singlet Cooper pairs in  CeCu$_2$Si$_2$, CeCoIn$_5$ and UPd$_2$Al$_3$.
On the other hand, no reduction in the Knight shift was observed in UPt$_3$ and UNi$_2$Al$_3$, although the estimated values of $\chi_{\gamma}$ and $\chi_{T_1}$ are large enough to be probed experimentally.
The odd-parity (spin triplet) superconductivity is therefore concluded in these compounds.
The Knight shift result provided  a convincing way to classify the HF superconductors into either even- or odd-parity pairing, as long as the system has Kramers degeneracy.

In this letter, we report the measurements of the $^{29}$Si Knight shift $^{29}K$ to address a possible order parameter symmetry of SC in the noncentrosymmetric superconductor CePt$_{3}$Si.
Recently, Ueda {\it et al.} have reported that no appreciable change is present in $K$  for $H\parallel c$-axis across $T_{c}$.
However, no detailed description is given on how to experimentally extract the spin part in $^{29}K$ from the $^{29}$Si NMR spectrum or how to estimate $\chi_{\rm s}$ from the available experimental data on this compound.\cite{CePt3Si_Ueda}
Here, we present the precise data of $^{29}K$ along and perpendicular to the tetragonal $c$-axis and show that its spin part does not change across $T_{c}$, irrespective of the crystal direction, although the estimated $\chi_{\rm s}$ should decrease even for a strong spin-orbit coupling regime if the spin-singlet pairing state were realized. 

Single crystals were successfully prepared by both the Bridgeman method and mineralization.\cite{CePt3Si_Hashimoto, CePt3Si_Yasuda}
The high quality of the sample is demonstrated by an observation of de Haas-van Alphen oscillations.
Furthermore, the estimated mean free path $l_{tr}=1200-2700$ \r{A} is one order of magnitude larger than the coherence length $\xi \simeq 100$ \r{A}, implying that CePt$_{3}$Si is in the clean limit.\cite{CePt3Si_Hashimoto, CePt3Si_Yasuda}
For the $^{29}$Si-NMR measurement, the samples are crushed into a coarse powder to make the applied rf-field penetrate easily.
The powder is partially oriented along the $c$-axis parallel to the magnetic field $H$ and at $T=4.2$ K. 
The NMR measurement was carried out using a conventional phase-coherent spectrometer and a $^{3}$He-$^{4}$He dilution refrigerator at a constant magnetic field.
The NMR spectrum was obtained using the Fourier transform technique of the spin-echo signal above $T_{\rm N}=2.2$ K, and integrating the spin-echo signal point by point as a function of frequency below $T_{\rm N}$.

\begin{figure}[thbp]
  \begin{center}
    \includegraphics[keepaspectratio=true,height=77mm]{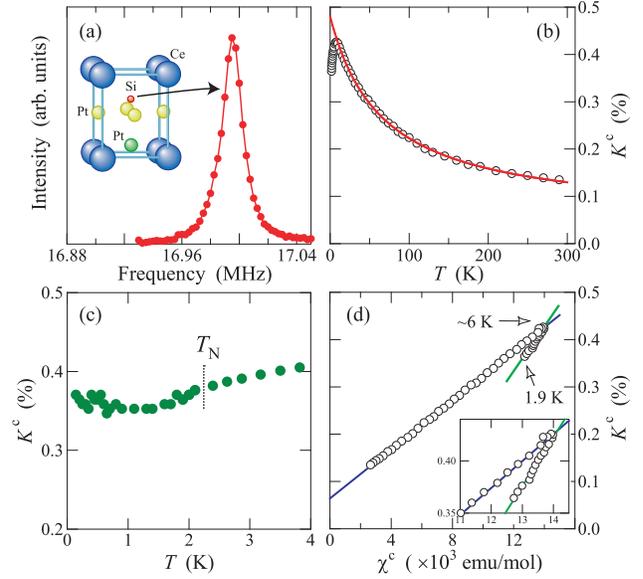}
  \end{center}
  \caption{(color online) (a) $^{29}$Si-NMR spectrum at $T=4.2$ K and $H=2.0012$ T. $T$ dependence of $K^c$ of $^{29}$Si along $c$-axis in (b) $T$=1.9 K - 300 K and (c) below 4 K. The solid line indicates a Curie-Weiss fit (see text). (d) $K^{c}$ vs $\chi^{c}$ plot with $T$ as implicit parameter. The solid lines indicate $K^{c}=(A_{\rm hf}^{c}/N_{\rm A}\mu_{\rm B})\chi^{c}$ with respective hyperfine coupling constants $A_{\rm hf}^{c} = 1.45$ and 2.63 kOe/$\mu_{\rm B}$ above and below 6 K.}
\end{figure}

Figure 1(a) shows the $^{29}$Si-NMR spectrum in which the full width at half maximum (FWHM) $\Delta f$ is as small as $\sim$ 17 kHz at 4.2 K, pointing to the high quality of the sample.
From the comparison with the spectral shape for unoriented samples, it is suggested that an anisotropy in the NMR spectrum is not so appreciable, indicating that the hyperfine interaction for $^{29}$Si nuclei is rather isotropic in contrast to the case for $^{195}$Pt.
Figures 1(b) and 1(c) show the $T$ dependence of $^{29}K$ along the $c$-axis ($K^{c}$), the value of which is scaled to the magnetic susceptibility $\chi_{\rm obs}$ following the Curie-Weiss law above 10 K.\cite{CePt3Si_Takeuchi}
In general, the Knight shift is expressed as $K(T)=K_{\rm s}(T)+K_{\rm orb}$.
Here, $K_{\rm s}$ and $K_{\rm orb}$ are the spin and orbital components of the Knight shift, respectively.
Note that $K_{\rm s}$ is responsible for the $T$ dependence of the shift.
To estimate $K_{\rm orb}$ and $K_{\rm s}$, $K^{c}$ is fitted with the relation $K^{c}=C/(T+\theta)+K_{\rm orb}$ as indicated by the solid line in Fig. 1(b) with $\theta=-66$ K and $K_{\rm orb}=0.05$\%.
Figure 1(d) shows the $K^{c}$ vs $\chi^{c}$ plot with $T$ as an implicit parameter, where $\chi^{c}$ is the magnetic susceptibility for $H\parallel c$-axis. Note that its slope changes below $\sim 6$ K, which is comparable to the energy separation 7 K  between the ground state CEF level and the first excited CEF level, suggesting that the hyperfine coupling process is mediated by the conduction electron spin polarization via hybridization with the $4f$ derived states.  From the relation $K^{c}=\frac{A_{\rm hf}^{c}}{N_{\rm A}{\mu_{\rm B}}}\chi^{c}$, hyperfine coupling constants $A_{\rm hf}^{c}=1.45$ and 2.63 kOe/$\mu_{\rm B}$ are estimated above and below 6 K, respectively. 
Furthermore, from an extrapolation of this line to a high-temperature limit, that is, the intercept with the vertical axis in Fig. 1(d), a tentative value of $K_{\rm orb}$ is estimated to be $\sim0.06$\%. 
%
\begin{figure}[thbp]
  \begin{center}
    \includegraphics[keepaspectratio=true,height=70mm]{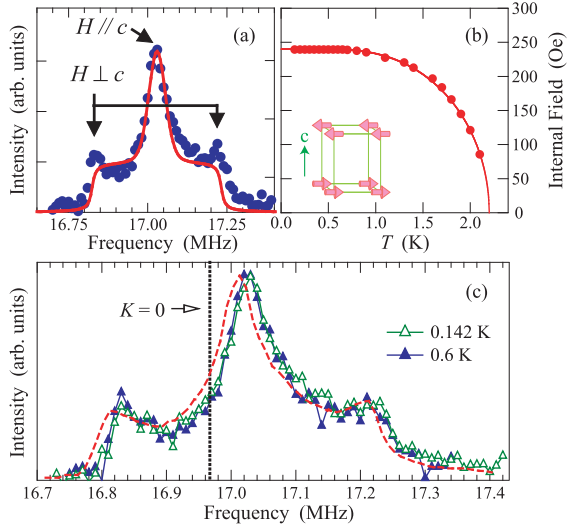}
  \end{center}
  \caption{(color online) (a) $^{29}$Si-NMR spectrum at $T=0.142$ K. The solid line indicates a simulation of the NMR spectrum for the partially aligned powder sample. Here, the internal fields at the Si site are deduced as $H_{{\rm int},\perp}=240$ Oe and $H_{{\rm int},c}=0$. (b) $T$ dependence of $H_{\rm int,\perp}$ below $T_{\rm N}$. The solid line is a guide to the eye. The inset shows the spin structure in the antiferromagnetic state.\cite{CePt3Si_Metoki} (c) $^{29}$Si-NMR spectra just above $T_{c}$ (closed triangle) and well below $T_{c}$ (open triangle) at $H=2.0061$ T. The spectrum illustrated by the broken line indicates a simulation assuming a $0.1 \%$ decrease in $K$.}
\end{figure}

Figure 2(a) shows the $^{29}$Si-NMR spectrum in the coexisting state of AFM and SC at $T=0.142$ K and $H=2.0061$ T.
In general, provided that an internal field $H_{\rm int}$ smaller than an external field is randomly directed for the powder sample in the AFM state, a rectangular shape for such a spectrum is expected with the width of $2H_{\rm int}$.
From the AFM structure, which is shown in the inset of Fig. 2(b),\cite{CePt3Si_Metoki} it is expected that $H_{{\rm int},c}$ along the $c$-axis is smaller than the $H_{{\rm int},\perp}$ in the plane. Accordingly, if the sample were oriented along the $c$-axis, a sharp spectral shape would be observed.  The observed unique spectrum suggests, however, that the powder sample is not perfectly aligned along the $c$-axis. Instead, by assuming the partial orientation along the $c$-axis and incorporating the anisotropic internal fields with $H_{{\rm int},\perp}=240$ Oe and $H_{{\rm int},c}=0$, the spectrum at $T=0.142$ K is well simulated by a solid line in Fig. 2(a). Here, it is anticipated that about $60$\% of the powder sample is oriented along the $c$-axis. From a central peak, the Knight shift $K^{c}$ for $H\parallel$ $c$-axis is deduced, whereas  both edges of the spectrum $H_{\rm high}$ and $H_{\rm low}$ are given by the respective relations $H_{\rm high}=H_{{\rm int},\perp}+K^{\perp}H$ and $H_{\rm low}=-H_{{\rm int},\perp}+K^{\perp}H$, allowing us to estimate  $K^{\perp}=(H_{\rm high}+H_{\rm low})/2H$. 
From the measurements of the $T$ dependence of this spectral shape, $H_{{\rm int},\perp}$,  $K^{\perp}$ and $K^{c}$ are extracted as a function of $T$.

Figure 2(b) shows the $T$ dependence of $H_{{\rm int},\perp}$ below $T_{\rm N}=2.2$ K and its saturation in the superconducting state below $T\sim 0.7$ K.  Therefore, $K^{\perp}$ and $K^{c}$ below $T_c$ are extracted from the $T$ dependence of the spectrum as shown in Fig. 2(c).
A remarkable finding is that there is no obvious change in $K^{\perp}$ and $K^{c}$ across $T_c\sim 0.52$ K in the range $T=0.142 - 0.6$ K, suggesting that the spin susceptibility seems to be unchanged across $T_{c}$ at $H= 2.0061$ T.

Figures 3(a) and 3(b) show the $T$ dependences of $K^{\perp}$ and $K^{c}$ along with Pt-NMR $(1/T_{1}T)$ in Fig. 3(c) \cite{CePt3Si_Yogi} to make sure that $T_{c}=0.52$ K under the magnetic field. Apparently, $K^{c}$ and $K^{\perp}$ do not decrease across $T_{c}=0.52$ K as denoted by the broken lines. 
\begin{figure}[thbp]
  \begin{center}
    \includegraphics[keepaspectratio=true,height=70mm]{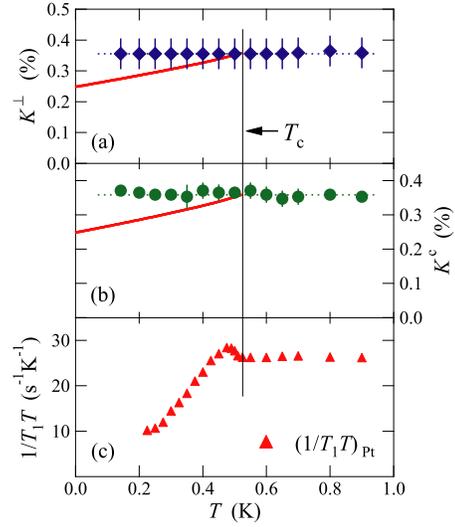}
  \end{center}
  \caption{(color online) $T$ dependence of (a) $K^{\perp}$ ($H\perp c$-axis) and (b) $K^{c}$ ($H\parallel c$-axis) at $H=2.0061$ T. The solid lines indicate a calculation obtained by assuming a 0.11\% decrease in $K_{\rm s}$ for spin-singlet SC with line nodes in a gap function. Here, the gap function is assumed to be $\Delta(\theta, \phi)=\Delta_{0} \cos\theta e^{i\phi }$ with $2\Delta_{0} /k_{\rm B}T_{c}=5$. (c) $T$ dependence of $^{195}$Pt $1/T_{1}T$ at $H\approx 2$ T.\cite{CePt3Si_Yogi}}
\end{figure}

Recently, it has been argued that $\chi_{\rm s}$ of heavy quasiparticles in the HF superconductors is reasonably estimated within the framework of the Fermi-liquid theory, focusing on the NMR and specific-heat results.\cite{NMR_Ks_Tou} 
In these analyses, the relation of $\chi_{\rm s}$ to the spin part in the Knight shift, $K_{\rm s}$, and the Sommerfeld coefficient, $\gamma$, is described as
\begin{equation}
K_{\rm s,\gamma}=\frac{A_{\rm hf}}{N_{\rm A}\mu_{\rm B}}\frac{\gamma \mu_{\rm eff}^2}{\pi^{2} k_{\rm B}^{2}}R,
\end{equation}
where $\mu_{\rm eff}$ is the effective moment and $R$ is the Wilson ratio.
From the values of $A_{\rm hf}=2.63$ kOe/$\mu_{\rm B}$, $\gamma_{s}=180$ mJ/(K$^{2}$mol),\cite{CePt3Si_Bauer} $\mu_{\rm eff}=2.54$ $\mu_{\rm B}$ \cite{CePt3Si_Takeuchi} and assuming that $R=1$, $K_{\rm s,\gamma}$ is estimated to be 0.25\%.
Here, we note that the influence of vortex cores should be taken into account particularly in the case of unconventional superconductors with nodes in a gap function for deducing the size of reduction in $K_{\rm s}$ under the magnetic field. The recent thermal-transport measurements have suggested that CePt$_3$Si is most likely to have line nodes.\cite{CePt3Si_Izawa}
In the presence of line nodes, where the density of states $N(E)$ at the Fermi level DOS has a linear energy dependence, $N(H)$ increases in proportion to $\sqrt{H}$ due to the Doppler shift in the energy of quasiparticles in a circulating superconducting flow around vortices.
Therefore, $K_{\rm s}$ could increase with $\sqrt{H}$. For instance, in the case of a $d$-wave with the line nodes, the predicted relation $N(H)/N_0=g\sqrt{H/H_{c2}}$ was confirmed with a prefactor $g$ in the range of $0.5-0.7$.\cite{Zheng} 
If this Doppler shift were incorporated, $K_{{\rm s}, \gamma}=0.11-0.15$\% would be evaluated at $H\simeq 2$ T.

The solid lines in Figs. 3(a) and 3(b) show a calculation obtained by assuming $K_{\rm s}(H)=0.11$\% at $H=2.0061$ T in the case of line-node spin-singlet SC with inversion symmetry.
Here, a gap function is assumed to be $\Delta(\theta, \phi)=\Delta_{0} \cos\theta e^{i\phi }$ with $2\Delta_{0} /k_{\rm B}T_{c}=5$ as discussed below. Both $K^c$ and $K^{\perp}$ seem to be unchanged across $T_{c}$; however, because the measured field $H=2.0061$ T is not significantly lower than the upper critical field $H_{c2}=3.2$ T, further experimentation at fields much lower than $H_{c2}$ is important to obtain a firm evidence for the invariance in $\chi_{\rm s}$ across $T_{c}$.
 \begin{figure}[thbp]
  \begin{center}
    \includegraphics[keepaspectratio=true,height=40mm]{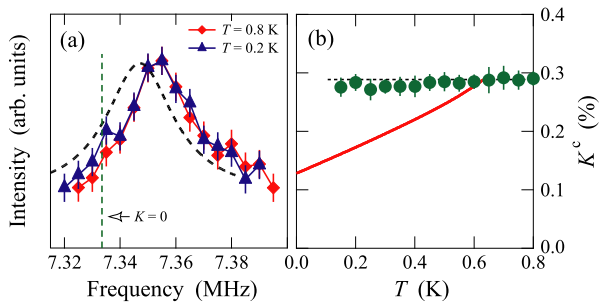}
  \end{center}
  \caption{(color online) (a) $^{29}$Si-NMR spectra above $T_{c}$ (closed square) and well below $T_{c}$ (closed triangle) at $H=0.8671$ T. The spectrum shown by the broken line indicates a simulation assuming a $0.1 \%$ decrease in $K$.  (b) $T$ dependence of $K^{c}$ ($H \parallel c$-axis). The solid lines indicate a calculation obtained by assuming a $K_{\rm s}=0.16$\% decrease in the case of spin-singlet SC with line nodes in a gap function. Here, the gap function is assumed to be $\Delta(\theta, \phi)=\Delta_{0} \cos\theta e^{i\phi }$ with $2\Delta_{0} /k_{\rm B}T_{c}=5$ (see text).}
\end{figure}

Figure 4(a) shows the NMR spectra at $T=0.2$ and 0.8 K at $H=0.8671$ T, which seem to reveal no appreciable change across $T_{c}=0.63$ K. A simulation of the spectrum, which is shown by the broken line in Fig. 4(a), assumes a $0.1$\% decrease in $K$. It would be expected that the present measurement could detect a $0.1$\% decrease in $K$ across $T_c$ if any.
A slight decrease in $K^{c}$ upon cooling is ascribed to the diamagnetic shift, which is estimated as $H_{\rm dia}\sim -0.4$ Oe using the relation $H_{\rm dia}=-H_{c1}\ln (\beta e^{-1/2}d/\xi)/\ln \kappa$.\cite{Book_SC}
Here, we used $H_{c1}\sim 1$ mT,\cite{CePt3Si_Bonalde} $\beta=0.381$ for the triangular lattice, $d\sim 516$ \r{A} and $\kappa \sim 140$.\cite{CePt3Si_Bauer}
$H_{\rm dia}$ yields the diamagnetic shift $K_{\rm dia}\ll 0.01$\%, which is negligibly small.
$K_{{\rm s}, \gamma}=0.16-0.19$\% estimated at $H=0.8671$ T becomes larger than $K_{{\rm s}, \gamma}=0.11-0.15$\% at $H=2$ T because the Doppler shift $N(H)$ is reduced. The solid line in Fig. 4(b) shows a calculation obtained by assuming a $K_{\rm s}=0.16$\% decrease in the case of spin-singlet SC with line nodes in a gap function. Here, a gap function is assumed to be $\Delta(\theta, \phi)=\Delta_{0} \cos\theta e^{i\phi }$ with $2\Delta_{0} /k_{\rm B}T_{c}=5$. 
It is clearly seen that $K^{c}$ does not change across $T_c$, which allows us to conclude that $\chi_{\rm s}(T)$ remains unchanged below $T_{c}$.

Frigeri {\it et al.} showed that stable superconducting phases would give a very weak change in $\chi_{\rm s}$ for the fields along the $c$-axis irrespective of the pairing symmetry.\cite{CePt3Si_Frigeri2}
On the other hand, $\chi_{\rm s}$ for the fields in the basal plane undergoes a moderate reduction, that is, $\chi_{\rm s}(0)\sim \chi_{\rm s}(T=T_c)/2$. Although the $K^{\perp}$ at $H=2$ T seems to be unchanged across $T_{c}$, more precise data are required at lower fields to provide convincing evidence that $\chi_{\rm s}$ in the basal plane decreases below $T_{c}$. In any case, however, taken together with the intimate result on the $1/T_1$ below $T_{c}$,\cite{CePt3Si_Yogi} the SC in CePt$_3$Si is characterized  by the two-component order parameter composed of spin-singlet and spin-triplet pairing components.

In conclusion, the $^{29}$Si Knight-shift of CePt$_{3}$Si showed no change across $T_{c}$ irrespective of field direction at $H=2.0061$ T and for $H\parallel c$-axis at $H=0.8671$ T.
It has been shown that the spin susceptibility, which is estimated to be sizable, remains unchanged for the fields along the $c$-axis. Taken together with the $T_{1}$ result that reveals the peak just below $T_{c}$, the superconductivity in CePt$_3$Si without an inversion center is reinforced to be in the novel Cooper pairing state with the two-component order parameter composed of spin-singlet and spin-triplet pairing components for large spin-orbit coupling.

We thank H. Tou and K. Ishida for valuable comments.
This work was supported in part by a Grant-in-Aid for Creative Scientific Research (15GS0213), MEXT and the 21st Century COE Program of the Japan Society for the Promotion of Science, and also by the Austrian FWF, P16370.

\end{document}